\documentclass[aps,prl,twocolumn,groupedaddress]{revtex4-2}

\usepackage{graphicx}
\usepackage{dcolumn}
\usepackage{bm}
\usepackage{hyperref}
\usepackage{siunitx}
\usepackage{amsmath, amsfonts}
\usepackage{ulem}

\begin{document}


\title{Critical percolation in the ordering kinetics of twisted nematic phases}


\author{R. A. L. Almeida}
\email[]{renan.almeida@ufrgs.br}
\affiliation{Instituto de F\'\i sica, Universidade Federal do
Rio Grande do Sul, CP 15051, 91501-970, Porto Alegre RS, Brazil}

\date{\today}

\begin{abstract}

I report on the experimental confirmation that critical percolation statistics underlie the ordering kinetics of twisted nematic phases in the Allen-Cahn universality class.
Soon after the ordering starts from a homogeneous disordered phase and proceeds towards a broken $\mathbb{Z}_2$-symmetry phase, the system seems to be attracted to the random percolation fixed point at a special timescale $t_{\textrm{p}}$. 
At this time, exact formulae for crossing probabilities in percolation theory agree with the corresponding probabilities in the experimental data.
The ensuing evolution for the number density of hull-enclosed areas is described by an exact expression derived from a percolation model endowed with curvature-driven interface motion.
Scaling relation for hull-enclosed areas versus perimeters reveals that the fractal percolation geometry is progressively morphed into a regular geometry up to the order of the classical coarsening length.
In view of its universality and experimental possibilities, the study opens a path for exploring percolation keystones in the realm of nonequilibrium, phase-ordering systems.
\end{abstract}


\maketitle


%

%
Phase-ordering kinetics or domain coarsening \cite{Bray02, Cugliandolo15} 
is a general phenomenon with universal features that plays key roles in 
solid alloys \cite{Allen79}, 
soft matter \cite{Tanaka00}, 
cell biology \cite{Boeynaems18},
bacterial populations \cite{McNally17},
and opinion models \cite{Dornic01}, 
to mention but a few examples beyond the classical spin model systems in condensed matter. 
Indeed, the most familiar and
%
%
iconic example is the ordering of ferromagnetic phases in the bidimensional kinetic Ising model after a quench from above to below the critical temperature.
%
Evolving with single flip kinetics, the mosaic of spin domains acquires a morphology statistically equivalent to that of critical percolation \cite{Arenzon07, Sicilia07, Barros09, Olejarz12} before developing the coarsening length, $R(t) \sim t^{1/2}$, that dynamical scaling hypothesis in this case relies upon \cite{Bray02, Cugliandolo15}; $t$ is the time elapsed from the quench.
%
At the continuum scaling limit, such a phenomenology is recast as a nonconserved scalar field evolved by the time-dependent Ginzburg-Landau equation (model A) with a symmetric double well potential with minima at $\pm \phi_0$ \cite{Bray02}.
The dynamics is concentrated at the motion of interfaces (i.e., the zeros of the scalar field), 
whose curvatures are reduced according to the Allen-Cahn (AC) equation \cite{Allen79}: $v = -D\kappa$, where $v$ is the normal velocity of an infinitesimal segment of the interface, $\kappa$ is the local curvature, and $D$ is a parameter.
%
%
%

Starting from a homogeneous disordered initial condition, 
the low-temperature dynamics of both discrete and continuum finite-size models quickly visit configurations characterized by the existence of giant percolating clusters, whose sizes, occupying a large fraction of the system, are at variance with the typical domain size kept at the microscopic level \cite{Blanchard14, Blanchard17, Azevedo22}. 
Because of the interplay between energy-conserving and energy-decreasing kinetical moves \cite{Azevedo22}, however, these initial percolating clusters are broken and rebuilt multiple times until the dynamics converges to the random percolation fixed point at a special timescale $t_{\textrm{p}}$ \cite{Blanchard14, Blanchard17}.
%
%
The underlying percolating structure is then permanently sealed at this time 
\cite{Blanchard14}, leaving the role of smoothing boundaries and coarsening domain areas for the asymptotic AC dynamics.
%
%
%
%
\begin{figure}[!t]
\includegraphics{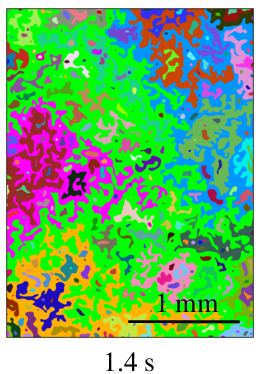}
\includegraphics{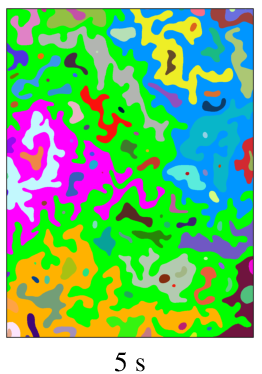}
\includegraphics{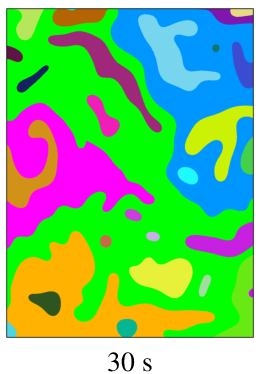}
\caption{\label{fig_mosaic}
(Color online) Mosaics of domains during the ordering of twisted nematic phases.
In a panel, each domain is artificially painted in a unique color. 
The 1st, 2nd, 3rd and 4th largest clusters are coherently painted in green, blue, orange, and magenta colors, respectively. 
Times elapsed from the quench are shown below each panel.
}\end{figure}

Interestingly, from $t_{\textrm{p}}$ onwards, the crossing probabilities for spin domains in the Ising-Glauber model, lying in a rectangle of aspect ratio $r$, numerically follow \cite{Barros09, Olejarz12} the probabilities exactly derived for critical percolation \cite{Watts96, Dubedat06}. %
%
%
As I shall show below, such a numerical result is here experimentally confirmed, along with the first observation of $t_{\textrm{p}}$ in real systems, in addition to the experimental confirmation for celebrated Cardy's formula \cite{Cardy92,Smirnov01}.
For free boundary conditions, a domain that crosses over a rectangle by means of a vertical spanning component, without having a horizontal spanning component, occurs with probability \cite{Watts96}
\begin{equation}
\mathcal{F}_{\overline{h}v}(r) = \frac{\eta(r)}{\Gamma(1/3)\Gamma(2/3)}
\textrm{ }_3F_2(1, 1, 4/3; 2, 5/3; \eta),
\label{eq_wattsformula_v}
\end{equation}
where $\Gamma(\cdot)$ and $\textrm{}_mF_n(a_1, ..., a_m; b_1, ..., b_n; \eta)$ are the Gamma and the generalized hypergeometric functions;
$\eta(r)$ is defined, and implicitly related to $r$, as
$\eta = [(1-k)/(1+k)]^2$, $r \in \mathbb{R}^+$, with $r = 2K(k^2)/K(1-k^2)$  \cite{Barros09}, where $K(u)$ is the complete elliptic integral of the first kind.
A $\pi/2$ rotation of the rectangle maps the horizontal direction onto the vertical direction, and vice-versa. Then, 
\begin{equation}
\mathcal{F}_{h\overline{v}}(r) = \mathcal{F}_{\overline{h}v}(1/r).
\label{eq_wattsformula_h}
\end{equation}
The probability for a dual-spanning configuration, $\mathcal{F}_{hv}$, is obtained from the normalization requirement,
\begin{equation}
2\mathcal{F}_{hv}(r) = 1 - \mathcal{F}_{h\overline{v}}(r) - \mathcal{F}_{\overline{h}v}(r),
\label{eq_wattsformula_hv}
\end{equation}
where the factor 2 enters because a configuration with no spanning cluster in critical percolation shall be, by the up-down symmetry, counted as a mosaic of dual-spanning type for the ordering problem \cite{Barros09}. 
The crossing probability for domains containing at least a vertical spanning component is Cardy's formula. It reads \cite{Cardy92, Smirnov01}:
\begin{equation}
\mathcal{F}_{hv}(r) + \mathcal{F}_{\overline{h}v}(r)
= \frac{3\Gamma(2/3)}{\Gamma(1/3)^2}\eta^{1/3}\textrm{}_2F_1(1/3, 2/3; 4/3; \eta)
\label{eq_cardyformula}
\end{equation}
%
%
%

By letting a critical percolation configuration, at the continuum scaling limit, evolve with the curvature-driven interface motion at zero temperature, one can also derive an exact expression for the number density of hulls with enclosed area between $A$ and $A+dA$, $n_{\textrm{h}}(A,t)dA$ \cite{Arenzon07},
\begin{equation}
n_{\textrm{h}}(A,t) = \frac{ 2c_{\textrm{h}} }{(A + \lambda_{\textrm{h}}t)^2},
\qquad c_{\textrm{h}} = \frac{1}{8\pi\sqrt{3}},
\label{eq_arenzonformula}
\end{equation}
with $c_{\textrm{h}}$ being a universal constant \cite{Cardy03}. 
Equation~\eqref{eq_arenzonformula} is valid for $A_0 \ll A \ll L^2$ and $t \geq t_{\textrm{p}}$, where $A_0$ and $L^2$ denote a microscopic area and the system area, respectively; 
$\lambda_{\textrm{h}}$ is a parameter. 
Despite being derived from a continuum model, Eq.~\eqref{eq_arenzonformula} describes the evolution of hull-enclosed areas in the kinetic Ising model with nonconserved dynamics evolving at low temperatures \cite{Arenzon07,Sicilia07}. 
A power-law decay consistent with $n_{\textrm{h}}(A) \sim A^{-2}$ was seen in experiment of chiral smectics \cite{Sicilia08}, a system different from the one presented below (see Ref.~\cite{Almeida21} and Supp. Mat. \cite{SM}). However, as far as Eq.~\eqref{eq_arenzonformula} is concerned, the universal value $2c_{\textrm{h}}$ -- hallmark of percolation statistics in coarsening dynamics \cite{Arenzon07} -- has not been observed yet.
%
%
%

In this Letter, I report on clear experimental evidence that critical percolation statistics underlie the ordering kinetics of twisted nematic phases in the Allen-Cahn universality class \cite{Bray02,Almeida21}. 
I do it so by confirming all the exact formulae, Eqs.~\eqref{eq_wattsformula_v} to \eqref{eq_arenzonformula}, based on percolation theory, including additional formulae and scaling relations derived from the interplay between ordering, percolation, and the coarsening regime. 
%
Unlike traditional studies with liquid crystals (that have focused on
the first-order, isotropic-mesophase thermal transition), 
here one explores the nontrivial electrohydrodynamic convection patterns (ECP) that arise in a class of nematic layers \cite{Kai89,*Kramer95}.
While keeping thermal effects finely controlled in the system, ordering kinetics 
can be electrically induced by genuine sudden transitions through second-order like points. 
The setup is also versatile: 
it allows inducing different types of ordering by transiting, e.g., between the nonequilibrium steady-states in the ECP regime.
This \textit{trium} of qualities (fast, high-controllable, versatile), associated with the convenient spatial and temporal scales of liquid crystalline systems, turns the experimental setup ideal for addressing universality issues in phase-ordering kinetics.
Below, I describe the specific experimental methods before presenting the results and discussions.
For more detail on the experimental setup, see Ref.~\cite{Almeida21} and Supp. Mat. \cite{SM}.
%
%
%
%

A twisted nematic liquid crystal (TNLC) cell was prepared by injecting a solution of
N-4-methoxybenzylidene-4-butylaniline
(purity  $> \qty{98}{\%}$) doped with $\qty{0.01}{wt\%}$ of tetrabutylammonium bromide in a rectangular region, $\qty{12}{\mu m}$ $\times$ $\qty{16}{m m} \times \qty{16}{m m}$, enclosed by parallel glass plates and polyester spacers. 
Inner surfaces of the plates, coated with indium tin oxide and polyvinyl alcohol, 
were mechanically rubbed to set an orientation for the nematic field right on them. 
These orientations were made orthogonal between the plates to induce left- and right-hand twisted nematic conformations along the bulk.
%
%
%

For optical observations, I inserted the cell on the stage of a IX73 Olympus microscope before illuminating it with circularly polarized green-filtered light.
Images formed by the light transmitted through the TNLC layer
were recorded by a B1620 Imperx camera. 
Each image comprises an area $L^2 = 1208a \times 1608a$ with pixel size $a = \qty{1.82}{\mu m}$.
The temperature of the TNLC layer was kept at $\qty{25}{^\circ C}$ with fluctuations of at most $\qty{10}{mK}$.
%
%
%

To induce ordering kinetics in the material, a sinusoidal voltage ($\qty{70}{V}$; $\qty{100}{Hz}$) was applied through the cell to generate the high density of string-like topological defects featuring the Dynamical Scattering Mode 2 (DSM2) \cite{Joetz86, Kai89}. 
DSM2 provides a nematic-disordered initial condition because the nematic order becomes chaotic with short correlations in space and time, nearly $\qty{1}{\mu m}$ and $\qty{10}{ms}$, respectively, for a $\qty{50}{\mu m}$ thick nematic layer under a.c. electric field ($\qty{60}{V}$; $\qty{150}{Hz}$) \cite{Joetz86}.
After letting the cell by \qty{2}{min} in DSM2, the field was suddenly removed -- definition of time $t = \qty{0}{s}$ -- 
and the stochastic ordering of twisted nematic phases was kept tracked. 
%
%
%

The ordering is quantified by a binary scalar nonconserved order parameter endowed with nearly-symmetric, AC dynamics; the typical domain size grows as $R(t) \sim t^{1/2}$ \cite{Almeida21}.
Measuring the shrinking rate of circular domains, the timescale of the curvature-driven motion can be quantified by $D = \lambda_{\textrm{h}}/2\pi = \qty{122(4)}{\mu m^2 s^{-1}}$ \cite{Almeida21}, from where one reads $\lambda_{\textrm{h}} = \qty{767(25)}{\mu m^2 s^{-1}}$. 
Using this setup, I focus on geometrical aspects of the domain morphology to test exact predictions based on percolation. 
To this aim, 1000 independent ordering histories lasting \qty{30}{s} each were collected. Images were acquired at $\qty{5}{s^{-1}}$ frame rate.
In the analysis, a domain is defined as a connected path of the same phase.
Each domain contains an external contour defined as its hull.
Domains and hull-enclosed areas were detected by a labelling \cite{Hoshen76} and a biased-walker algorithm \cite{Sicilia07}.
The hull perimeter is defined as the number of broken bonds of each pixel at the hull times the pixel size.
%
%

By following the evolution of the domain morphology in Fig.~\ref{fig_mosaic}, 
we observe that the macroscopic shapes of largest domains in the panel at $\qty{1.4}{s}$ are preserved by the dynamics up to, at least, the latter panel at $\qty{30}{s}$.
%
The main changes during the evolution of these domains take place at their interfaces and areas:
the former becomes smoother because of the curvature-driven motion, and the last increases as a result of the shrinking and disappearance of inner domains.

For $t \geq \qty{1.4}{s}$, the largest domain in Fig.~\ref{fig_mosaic} crosses over the opposite sides of the image by spanning it along both of its horizontal and vertical directions.
%
Crossing events like this are the rule since all 
analysed mosaics
%
have at least a domain that crosses over the image.
%
%
%
Both the crossing event and the crossing type, however, vary upon the geometry considered.
To quantify these events over a simple geometry, I consider a rectangle of aspect ratio $r = l_x / 1208a$, $r = 0.2, 0.3, ..., 1$, located over the original image.
The rectangle is oriented such that its $x$ and $y$ Cartesian axes are parallel to the short and long sides of the image, respectively; its upper left corner is fixed at that same corner of the panels.
Since images were taken in a region far from the borders of the sample, it is checked that different positions of the rectangle does not alter the results.
%
%
%
\begin{figure}[!t]
\includegraphics{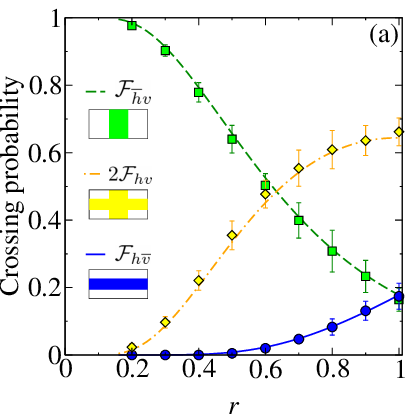}\quad
\includegraphics{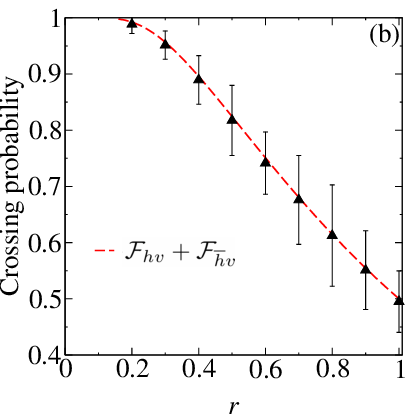}
\caption{\label{fig_cross}
Exact crossing probabilities for percolation theory (lines), as a function of the aspect ratio of a rectangular arena, describe experimental data for ordering kinetics (symbols). 
(a) Probabilities for only vertical ($\square$), only horizontal ($\circ$), and dual-spanning ($\diamondsuit$) configurations. 
(b) Cardy's formula -- right-hand side of Eq.~\eqref{eq_cardyformula} -- compared with experimental crossing configurations containing a vertical component.
Uncertainties correspond to 1 standard deviation in (a), and to the error propagation of standard deviations through the left-hand side of Eq.~\eqref{eq_cardyformula} in (b). 
Measurements taken at \qty{1.4}{s}.}
\end{figure}

Figure~\ref{fig_cross}(a) shows the experimental crossing probabilities for each one of the three crossing types, as function of $r$, computed from configurations at $\qty{1.4}{s}$.
Ordering times within $\qty{1.4}{s} \leq t \leq \qty{4}{s}$ yield statistically similar results.
The outcomes are to be compared with the exact results for percolation theory, Eqs.~\eqref{eq_wattsformula_v}, \eqref{eq_wattsformula_h}, \eqref{eq_wattsformula_hv}, shown as dashed or solid lines in the plot.
%
Notice that the three exact curves have monotonic behaviors easily distinguishable, one from another.
%
While $\mathcal{F}_{\overline{h}v}$ quickly decreases from 1 (for the thin slab geometry at $r < 0.2$) to $\approx 0.18$ (for the square geometry at $r = 1$), 
$2\mathcal{F}_{hv}$ varies in an opposite trend, augmenting from 0 to $\approx 0.64$ along the increasing scale for $r$.
%
In its turn, being smaller or at most equal than their counterparts, $\mathcal{F}_{h\overline{v}}$ has only a moderate lift with $r$ from 0 until the meeting point $\mathcal{F}_{\overline{h}v}(1) = \mathcal{F}_{h\overline{v}}(1)$. 
%
By noting the exact result $\mathcal{F}_{hv}(1) = 1/4 + (\sqrt{3}/4\pi)\ln(27/16) = 0.322...$ \cite{Maier03},
we can read $2\mathcal{F}_{hv}(1) = 0.644...$. 
Using this value in Eq.~\eqref{eq_wattsformula_hv}, we find $\mathcal{F}_{h\overline{v}}(1) = 1/2 - \mathcal{F}_{hv}(1) = 0.177...$ \cite{Barros09}.

Over the whole range of $r$, the experimental results are well described by the exact formulae for critical percolation -- for all the three crossing types.
%
The most likely values for the probabilities $\overline{h}v$ and $hv$ are in excellent agreement with Eqs.~\eqref{eq_wattsformula_v} and \eqref{eq_wattsformula_hv}, correspondingly. Their uncertainties are relatively small. Given that the measures are realized on a partial region of the sample, and that the rectangle defining a crossing event is considerably smaller than the image, such an agreement is yet more impressive.
%
The data for the $h\overline{v}$ crossing type is right on the top of $\mathcal{F}_{h\overline{v}}(r)$. 
%
For the special square geometry, probabilities for the $\overline{h}v$ and $h\overline{v}$ types are statistically equal to $0.169(37)$, a value that encompasses $\mathcal{F}_{\overline{h}v}(1) = 0.177...$. By consistence, the dual crossing probability in the liquid crystal setup is $0.66(4)$ at $r = 1$, again in agreement with the prediction $2\mathcal{F}_{hv}(1)=0.644...$. Finally, for the meeting point $\mathcal{F}_{\overline{h}v} = 2\mathcal{F}_{hv} \approx 0.48$ at $r \approx 0.63$, the closest experimental data available gives $0.49(4)$ at $r = 0.6$.

%
Having seen the accord with percolation solutions for the fundamental triad of crossing probabilities, the measurements for $\overline{h}v$ and $h\overline{v}$ crossing types can also be combined to confirm Eq.~\eqref{eq_cardyformula}, Cardy's formula, shown in Fig.~\ref{fig_cross}(b).
All of this accord, however, is only reached for configurations \textit{from} $\qty{1.4}{s}$, a fact that unveils $t_\textrm{p} = \qty{1.4(1)}{s}$ for a square of side $l_x \sim 10^3a$.
%
%

\begin{figure}[!b]
\includegraphics{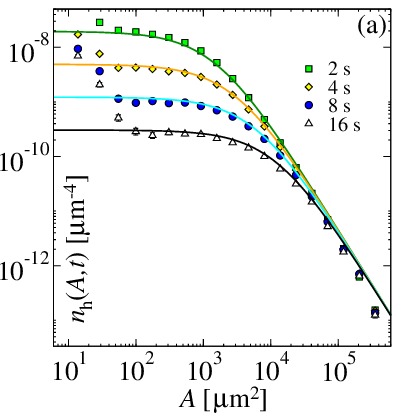}\qquad
\includegraphics{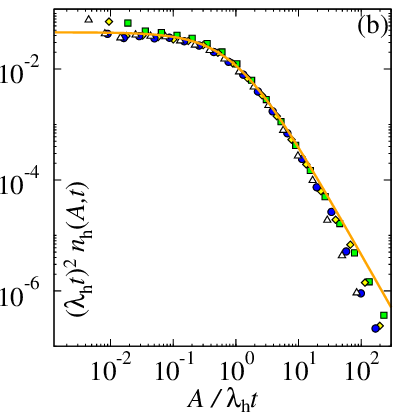}
\caption{\label{fig_hullarea} 
(a) Number density of hull-enclosed areas in the liquid crystal data (symbols) at several times compared with the corresponding proposed formula (lines), Eq.~\eqref{eq_arenzonformula} with $\lambda_\textrm{h} = \qty{767}{\mu m^{2}s^{-1}}$. 
(b) Master scaling form of data in (a). 
Error bars are one standard deviation of the mean. }\end{figure}

Now we turn to quantify the evolution of the domain morphology.
%
Figure~\ref{fig_hullarea}(a) shows results for $n_{\textrm{h}}(A,t)$ after exclusion of domains that touch a border of the image.
%
At fixed $t$, $n_{\textrm{h}}(A,t)$ is formed by three parts along the $A$ axis. 
%
In the smallest area part, $\qty{10}{\mu m^2} < A < \qty{50}{\mu m^2} \approx 27a$, 
$n_{\textrm{h}}(A,t)$ probes tiny bubble-like clusters in addition to thermal domains that are not related to the coarsening dynamics \cite{Sicilia07}. 
Because of thermal domains, $n_{\textrm{h}}$ manifests a temperature-dependent decreasing with $A$ that numerically can be accounted for by equilibrium distributions \cite{Sicilia07}.  
%
After such an initial decreasing with $A$, $n_{\textrm{h}}$ has a plateau region.
The extension of this plateau is delimited by the time-dependent coarsening area, $R^2(t) \sim t$.
Because of the curvature-driven motion, small domains at this regime shrink and disappear first than those having unusually large sizes.
As a result, $n_{\textrm{h}}(A,t)$ plateau shifts down as time elapses. 
Unlike in \cite{Sicilia08}, this temporal dependence is here clearly observed. 
%
Residing on large areas, on the other hand, are the structures similar to critical percolation clusters \cite{Blanchard14,Blanchard17}. 
After $t_\textrm{p}$, relaxation of these large structures becomes much slower than that of typical domains, so that 
$n_{\textrm{h}}(A,t)$ power-law decay is essentially time-independent. 
%

In Fig.~\ref{fig_hullarea}(a), I also plot the exact predictions from Eq.~\eqref{eq_arenzonformula} using the most likely value for the experimental settings (assumed hereafter), $\lambda_{\textrm{h}} = \qty{767}{\mu m^2 s^{-1}}$.
%
Remarkable agreement with theory is seen for both the plateau and the power-law regimes, $27a \ll A \ll L^2 \sim 10^6$, covering nearly one decade of variation in time. 
Minor deviations at small and large areas are due to the finite thermal length and system size, respectively \cite{Arenzon07,Sicilia07}.
%
In its master and universal form, Fig.~\ref{fig_hullarea}(b),
the results respect the dynamical scaling hypothesis over the full spatial and temporal ranges. 
Noteworthy, the plateau's level is compatible with $2c_{\textrm{h}} = 0.0459...$, thus passing through the stringent test of Eq.~\eqref{eq_arenzonformula} -- see \cite{Arenzon07}. 
The typical area, $A \sim \lambda_{\textrm{h}}t$, demarcates the crossover to the power law inherited from the universal percolation statistics, $n_{\textrm{h}}(A,t) \sim A^{-2}$ at $\lambda_{\textrm{h}}t \ll A \ll L^2$.
%
%
%

\begin{figure}[!t]
\includegraphics{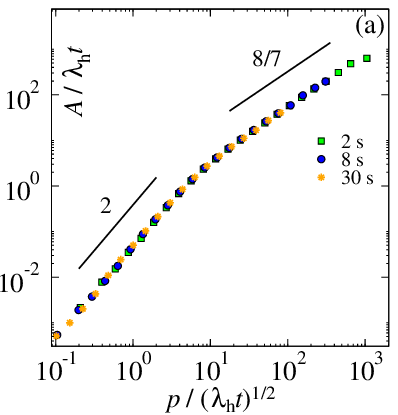}\quad
\includegraphics{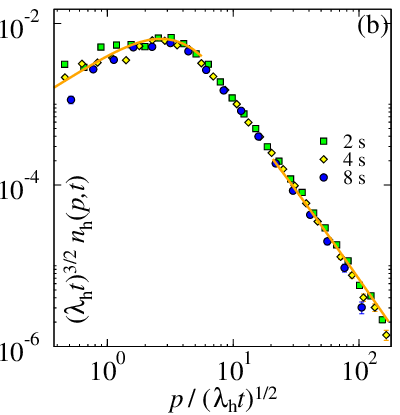}
\caption{\label{fig_hullperimeter} 
(a) Hull-enclosed areas versus associated perimeters in a master scaling form, Eq.~\eqref{eq_axp}, after bin average. 
Solid lines are guide to eyes and indicate slopes expected for regular (2) and percolation-like ($8/7$) geometries.
(b) Number density of hull perimeters formed by twisted nematic phases (symbols) compared with exact predictions (lines), Eq.~\eqref{eq_siciliaformula}, far from the crossover $p/(\lambda_\textrm{h} t)^{1/2} \approx 7$. 
In both plots, $\lambda_{\textrm{h}} = \qty{767}{\mu m^2 s^{-1}}$. Error bars are one standard deviation of the mean.
}\end{figure}

The morphing of clusters into regular (i.e., non-fractal) structures due to the ensuing ordering can be observed through
a simple relation for hull-enclosed areas and associated perimeters \cite{Sicilia07},
%
\begin{equation}
\frac{ A }{\lambda_\textrm{h} t} \simeq b \bigg(\frac{p}{\sqrt{\lambda_\textrm{h} t}}\bigg)^\alpha.
\label{eq_axp}
\end{equation}
The typical length, $\sqrt{\lambda_h t}$, is used as a normalization factor; $b$ is a parameter, while $\alpha = 2$ for regular hull geometry, but $\alpha < 2$ for fractal hull geometry.
%
Figure~\ref{fig_hullperimeter}(a) shows the experimental outcomes for the 
pairs of hull-enclosed area versus associated perimeter, after bin average, in their dynamical scaling form, Eq.~\eqref{eq_axp}. 
Domains that touch a border of the image are excluded from the statistics.
%
The data indeed collapse onto a master function made of two power laws, 
$y \sim x^{\alpha}$, with $y = A /\lambda_{\textrm{h}} t$ and $x = p/\sqrt{\lambda_\textrm{h} t}$: 
one power law below, and the other above, the crossover scale $x_{\textrm{c}} \approx 7$.
%
To quantify them, 
the local slopes $\alpha_{\textrm{loc}}(x) = d(\ln y)/d(\ln x)$ from the master curves are averaged over the following regions to find:
$\alpha = 2.04(20)$ over $0.3 < x < 5$; 
and $\alpha = 1.16(10)$ over $20  < x < 200$ (uncertainties are one standard deviation of the local slopes). Amplitudes are $b \approx 0.046$ and $0.25$, respectively.
%
Note that $\alpha = 1.16(10)$ agrees with the exact value for percolation hulls, $\alpha = 8/7 = 1.142...$, 
obtained from $\alpha = 2/d_{\textrm{f}}$ \cite{note1} with the hull fractal 
dimension $d_{\textrm{f}} = 7/4$ \cite{Saleur87}.
%
Therefore, the fractal geometry of hulls in the data is progressively morphed into a regular geometry, $\alpha = 2$. This occurs because of the spreading of correlations set up to the order of the coarsening length $R(t) \sim \sqrt{\lambda_h t}$: 
interfaces are smooth up to such length, while larger boundaries, keeping the memory of the critical percolation state, are largely rough.

%
%
To conclude the analysis, we also studied the number density of hulls with perimeters between $p$ and $p+dp$, $n_{\textrm{h}}(p,t)dp$.
%
%
Using $n_{\textrm{h}}(A,t)$ from Eq.~\eqref{eq_arenzonformula}, and $A(p,t)$ from Eq.~\eqref{eq_axp}, 
one can derive the exact expression proposed in \cite{Sicilia07}:
\begin{equation}
(\lambda_{\textrm{h}} t)^{3/2}n_{\textrm{h}}(x) \simeq 
2\alpha b c_{\textrm{h}}\frac{ x(p,t)^{\alpha-1} }
{ (1 + b x^{\alpha})^2 },
\label{eq_siciliaformula}
\end{equation}
for $x(p,t)$ far from $x_{\textrm{c}}$. 
%
Equation~\eqref{eq_siciliaformula} describes hull perimeters arising in the kinetic Ising model after a quench from infinite to zero temperature \cite{Sicilia07}.  
%

Figure~\ref{fig_hullperimeter}(b) shows $n_{\textrm{h}}(p,t)$ computed in the twisted nematic setup. 
The plot displays the data in the collapsed, dynamical scaling form $f(x) = (\lambda_{\textrm{h}} t)^{3/2}n_{\textrm{h}}(x)$.
%
Aside from a temperature-dependent region related to thermal domains at $0.03 \leq x \leq 0.3$ (not shown), 
the universal scaling function $f(x)$ comprises two parts along the $x$ axis, from $x = 0.3$ onwards:
a smooth increase that ends at a local maximum $f \approx 0.006$ for $x^*\approx 3$,
and a power law decay, $f(x) \sim x^{-(\alpha+1)}$, in the tail $x \gg x_{\textrm{c}}$.
%
Both regimes are described by Eq.~\eqref{eq_siciliaformula}.
The formulae are consistently generated with values extracted from the analysis of Fig.~\ref{fig_hullperimeter}(a).
Explicitly, we use $b = 0.046$ and $\alpha=2$ for small scales, $0.3 < x < 5$; 
while $b = 0.25$ and $\alpha = 1.25$ (best result) for large scales $20 < x < 200$.
Theoretical curves, shown as solid lines in Fig.~\ref{fig_hullperimeter}(b),
agree with the experimental results over their full extension of validity, 
which happens far from $x_{\textrm{c}} \approx 7$.
%

In conclusion, the ordering kinetics of twisted nematic phases in the Allen-Cahn universality class, starting from a homogeneous disordered initial condition, acquires a domain morphology statistically equivalent to that of the critical percolation model soon after the ordering begins.
%
On theoretical grounds, this connection has allowed theoretical physicists to propose a set of exact formulae for the class of bidimensional nonequilibrium systems with a nonconserved scalar field.
%
As we have seen, many of these formulae are here experimentally confirmed: 
$(i)$ the crossing probabilities formulae for rectangular geometries \cite{Cardy92, Smirnov01, Watts96, Dubedat06}; 
$(ii)$ the evolution for the number density of hull-enclosed areas \cite{Arenzon07};
$(iii)$ and the evolution for the number density of hull perimeters \cite{Sicilia07}.
In addition, I also have observed $(iv)$ the existence of $t_\textrm{p}$ in a real system; 
besides measuring that $(v)$ the fractal percolation geometry is progressively morphed into a regular geometry along with the spreading of correlations -- the crossover from regular to irregular shapes occurring at the order of the coarsening length.
%
%
The fact that exact solutions proposed for simple models work well in a complex and real system is a far from trivial result in view of their conceptual and microscopic differences. The observed agreements, conversely, elegantly exemplifies the powerful concept of universality built on the pillars of symmetries, conservation laws and dimensionality.
Thermal effects below the clearing point of the liquid crystal are expected to merely renormalize nonuniversal constants yet preserving exponents and master scaling functions \cite{Bray02}. Unusual dynamics should arise when the final applied voltage is set in the regime of electrohydrodynamic convection.
%
%
%

Other important aspects of percolation \cite{Cardy98,Schramm01},
including the connection to Schramm-Loewner evolution (SLE$_6$) \cite{Schramm00}, are appealing directions for assessment in nonequilibrium systems.
%
The emergence of critical percolation in coarsening phenomena implies a new exponent $t_{\textrm{p}} \sim L^{ z_{\textrm{p}} }$ \cite{Blanchard14, Blanchard17, Azevedo22} to appear in the dynamical scaling hypothesis. 
A measurement of the exponent $z_{\textrm{p}}$ in continuum models or 
real systems is a key piece required to complete the picture. 
The early fluctuating formation and reshaping of percolating clusters, as well as the universal behavior for the dynamical cluster size heterogeneity at this regime \cite{Azevedo20, Mazzarisi21}, also constitutes an important path for future research.
%
%
Observations of critical percolation even in coarsening models containing quenched disorders \cite{Arenzon07}, 
conserved dynamics \cite{Sicilia09}, and long range interactions \cite{Agrawal22}, indicate the possible generality of the phenomenon here experimentally reported.

\begin{acknowledgments}
I am grateful to J. J. Arenzon for the motivation and discussions; to K. A. Takeuchi
for advice on experiments and early discussions.
%
This research was funded by KAKENHI from JSPS Grant No. JP16J06923, by the Brazilian National Council for Scientific and Technological Development – CNPq, and the Fundação de Amparo à Pesquisa do Estado do Rio Grande do Sul – FAPERGS, Grant No. 23/2551-0000154-3. Data are available upon reasonable request.
\end{acknowledgments}


\providecommand{\noopsort}[1]{}\providecommand{\singleletter}[1]{#1}%

\end{document}